%% file: main.tex
\documentclass[12pt]{article}
\usepackage[utf8]{inputenc}
\usepackage{comment,times}
\usepackage{xurl}
\usepackage{hyperref}
\usepackage{graphicx}
\usepackage[table,xcdraw]{xcolor}
\usepackage[caption=false]{subfig}
\usepackage{caption}
\usepackage{enumitem}
\usepackage{float}
\usepackage{longtable}
\usepackage{setspace}
\usepackage[skip=20pt plus1pt, indent=20pt]{parskip}
\usepackage{atbegshi}
\usepackage{booktabs}

\usepackage{algorithm}
\usepackage[noend]{algpseudocode}

\let\oldReturn\Return
\renewcommand{\Return}{\State\oldReturn}

\algrenewcommand\alglinenumber[1]{
  {\sf\footnotesize\addfontfeatures{Colour=888888,Numbers=Monospaced}#1}
}
\algrenewcommand\algorithmicrequire{\textbf{Precondition:}}
\algrenewcommand\algorithmicensure{\textbf{Postcondition:}}

\title{
  Selecting a suitable Parallel Label-propagation based algorithm for Disjoint Community Detection
  \vspace{-2ex}
}
\author{
  Subhajit Sahu \\
  Center for Security, Theory, and Algorithmic Research (CSTAR) \\
  IIIT Hyderabad, India - 500 032 \\
  subhajit.sahu@research.iiit.ac.in
}
\date{}

\begin{document}
\maketitle
% \tableofcontents
% \onehalfspacing

\begin{abstract}
Community detection is an essential task in network analysis as it helps identify groups and patterns within a network. High-speed community detection algorithms are necessary to analyze large-scale networks in a reasonable amount of time. Researchers have made significant contributions in the development of high-speed community detection algorithms, particularly in the area of label-propagation based disjoint community detection. These algorithms have been proven to be highly effective in analyzing large-scale networks in a reasonable amount of time. However, it is important to evaluate the performance and accuracy of these existing methods to determine which algorithm is best suited for a particular type of network and specific research problem. In this report, we investigate the RAK, COPRA, and SLPA, three label-propagation-based static community discovery techniques. We pay close attention to each algorithm's minute details as we implement both its single-threaded and multi-threaded OpenMP-based variants, making any necessary adjustments or optimizations and obtaining the right parameter values. The RAK algorithm is found to perform well with a tolerance of 0.05 and OpenMP-based strict RAK with 12 threads was 6.75x faster than the sequential non-strict RAK. The COPRA algorithm works well with a single label for road networks and max labels of 4-16 for other classes of graphs. The SLPA algorithm performs well with increasing memory size, but overall doesn't offer a favourable return on investment. The RAK algorithm is recommended for label-propagation based disjoint community detection.
\end{abstract}

\section{Introduction}
\input{01-introduction}

\section{Literature survey}
\input{02-survey}
\section{Evaluation}
\input{03-evaluation}

\section{Conclusion}
\input{04-conclusion}

\appendix
\small
\bibliographystyle{IEEEtran}
\bibliography{main}
\end{document}

%% file: 01-introduction.tex
There has been an unprecedented growth of real-world interconnected data \cite{graph-sakr21} that is naturally represented as graphs. These graphs emerging from real-world applications such as social and biological networks are massive in size leading to the adoption of parallelism as a way to address the scale issue. In addition to being massive, most real-world graphs are dynamic in nature with new edges being deleted and inserted all the time \cite{com-zarayeneh21}. Therefore, parallel algorithms for updating graph analytics over dynamic graphs are gaining significant research attention in recent years. Examples include the dynamic computation of centrality scores \cite{cent-lerman10, cent-li10, cent-bergamini16, cent-nathan17, cent-regunta21}, dynamic maintenance of biconnected components \cite{conn-galil91, conn-liang01, conn-haryan22}, dynamic computation of shortest paths \cite{path-narvaez00, path-chan08, path-khanda21}.
% There has been an unprecedented growth of interconnected data \cite{graph-sakr21}. In addition to being massive, most real-world graphs are dynamic in nature, with new edges being deleted and inserted all the time \cite{com-zarayeneh21}. Study of dynamic community detection approaches is thus an important area of research, where the objectives are not only to accelerate the process of discovering communities, but also to track their evolution over time.
% There has been an unprecedented growth of real-world interconnected data \cite{graph-sakr21} that is naturally represented as graphs. These graphs emerging from real-world applications such as social and biological networks are massive in size leading to the adoption of parallelism as a way to address the scale issue. In addition to being massive, most real-world graphs are dynamic in nature with new edges being deleted and inserted all the time \cite{com-zarayeneh21}. Therefore, parallel algorithms for updating graph analytics over dynamic graphs are gaining significant research attention in recent years. Examples include the dynamic computation of cnetrality scores \cite{ics2020,ipdps23}, dynamic maintenance of biconnected components \cite{ipdps22}, dynamic computation of shortest paths \cite{bhowmick}.

Community detection is a popular graph problem with practical applications in fields such as e-commerce, communication networks, and healthcare. It involves identifying groups of vertices in a graph, known as communities, that correspond to functional units in the system and are densely connected within the community but sparsely connected to the rest of the network. When we can identify these structures based on the topology of the network alone, they are said to be intrinsic (studied here). On the other hand, extrinsic communities are defined based on external information or attributes of the nodes, such as membership in a particular organization or geographic location. There are various types of communities that can be identified in a graph. Disjoint communities (studied here) are those in which each vertex belongs to exactly one community. Alternatively, overlapping communities allow each vertex to belong to more than one community \cite{com-gregory10}, while hierarchical communities have a multi-level membership structure.

The problem of community detection is NP-hard, and unlike graph partitioning, the number and size distribution of communities is not known beforehand. There are various methods for community detection, including label propagation \cite{com-raghavan07, com-gregory10, com-xie11}, random walk \cite{com-rosvall08}, diffusion \cite{com-kloster14}, spin dynamics \cite{com-reichardt06}, fitness metric optimization \cite{com-newman06, com-fortunato10}, statistical inference \cite{com-come15, com-newman16}, simulated annealing \cite{com-guimera05, com-reichardt06}, clique percolation \cite{com-derenyi05, com-maity14, com-gupta22}, and more. These methods can be divided into two main groups: divisive and agglomerative. Divisive methods, also known as top-down methods, start with all the vertices in a graph being part of a single community and iteratively identify and remove bridges to split into more well-connected communities \cite{com-girvan02, com-souravlas21}. Agglomerative methods, or bottom-up methods, merge two or more communities together such that a certain score is maximized \cite{com-zakrzewska15, com-zarayeneh21}. In certain cases, we have a set of relevant seed vertices of interest which are expanded to form communities surrounding them. This is known as seed set expansion \cite{com-whang13}.
% The problem of community detection is shown to be``\verb|NP-hard|''. Moreover,  unlike the problem of graph partitioning, neither the number of communities nor their size distribution is known a priori \cite{com-halappanavar17}. Nevertheless, there are several heuristic methods to obtain the community structure of a graph in the standard sequential setting \cite{}, the parallel setting \cite{}, and the distributed setting \cite{}. \kk{exist to obtain solutions that are good enough for most practical purposes \cite{com-brandes07, com-shi09}}.

%% kk: Why are we citing the ones listed at the end of the first two sentences?
Different community detection methods can return different communities, which can be evaluated using a quality function. Modularity is a popular fitness function introduced by Newman and Girvan which compares the number of intra-community edges to the expected number in a random-null model. It lies between -0.5 (non-modular clustering) and 1.0 (fully modular clustering). Optimizing this function theoretically results in the best possible grouping. Conductance is another fitness score that measures the community cut or inter-community edges. Constant Potts Model (CPM) is another quality function that overcomes some limitations of modularity \cite{com-traag11}.
% Different community detection methods can return different communities; these algorithms are heuristic-based. We can assess the quality of the communities obtained either through a comparison with ground-truth communities (if available), or through the use of quality metric using a fitness function. Modularity is a fitness function, introduced by Newman and Girvan, which compares the number of intra-community edges to the expected number in a random-null model. It lies between -0.5 (non-modular clustering) and 1.0 (fully modular clustering). Optimizing this theoretically results in the best possible grouping. The problem of community detection is then reduced to the problem of modularity maximization which is NP-complete. Conductance is another popular fitness score that measures the community cut or inter-community edges. An alternative quality function is the Constant Potts Model (CPM), which overcomes some limitations of modularity \cite{com-traag11}.

%% file: 02-survey.tex
There are several techniques that have been developed for detecting communities in networks. A number of them are based on modularity-optimization, hierarchical clustering, label propagation, region density, core clustering \cite{com-ruan15}, game theory, information theory (infomap) \cite{com-zeng19, com-zeng18, com-faysal19, infomap-rosvall09, com-rita20}, and biological evolution (genetics) \cite{com-taufan20, com-ghoshal19, com-lu20}. Metrics such as the modularity score \cite{com-newman06, com-blondel08, com-ghoshal19}, Normalized Mutual Information index (NMI) \cite{com-jain17, com-chopade17}, and Jaccard Index \cite{com-jain17} are used to compare the quality of communities obtained with different approaches.

The \textit{Raghavan Albert Kumara (RAK)}, also commonly known as the \textit{Label Propagation Algorithm (LPA)}, is a method used for identifying communities or groups within a network by initializing each vertex with a unique label and diffusing these labels across the graph. LPA is faster and more scalable than the Louvain algorithm, as it can be parallelized easily and does not require repeated optimization steps. However, it may not be as effective at identifying high-quality community structure in some types of networks and tends to produce communities with lower modularity than the Louvain algorithm. Additionally, the resulting solution produced by LPA may not be unique and may contain monster communities \cite{com-newman04, com-raghavan07}.
% The \textbf{Raghavan Albert Kumara (RAK)}, also commonly known as the \textit{Label Propagation Algorithm (LPA)}, is a popular technique for identifying communities or groups within a network. It works by initializing each vertex with a unique label, and then diffusing these labels across the graph. As labels are diffused, well-connected groups quickly reach a common label and expand until it is no longer possible.
% Label-propagation based methods, such as the RAK, can be faster and more scalable that the Louvain algorithm, as they do not require repeated optimization steps and can be parallelized more easily. While they are more robust to the initial order of the nodes, the resulting solution may not be unique \cite{com-newman04, com-raghavan07}. Can more easily identify densely interconnected communities. However, they tend to produce communities with lower modularity than the Louvain algorithm, and may not be as effective at identifying high-quality community structure in some types of networks. LPA can produce an aggregate of acceptable solutions, but the resulting solution many to be unique (unstable) \cite{com-newman04, com-raghavan07}, and may contain monster communities.

Improvements upon the LPA include the Node Influence Based Label Propagation Algorithm (NIBLPA) while provides a stable (non-random) mechanism of label choosing in the case of multiple best labels \cite{com-xing14}; Label influence Policy for Label Propagation Algorithm (LP-LPA) and Spreading Activation Label Propagation Algorithm (SALPA) \cite{com-berahmand18, com-sattari18} which address the issue of monster communities; the Three Stage (TS) algorithm which first identifies central nodes and combines communities for improved modularity (no information has been provided on its execution time) \cite{com-you20}; and Lie et al.'s approach which uses frontiers with alternating push-pull to reduce the number of edges visited and improve solution quality \cite{com-liu20}. Existing approaches for finding overlapping communities, such as the Community OVerlap PRopagation Algorithm (COPRA), or the Speaker-Listener Label Propagation Algorithm (SLPA), may also be used for disjoint communities by choosing the best label associated with each vertex \cite{com-gregory10, com-xie11}. While these modifications to LPA achieve communities of better quality, they can be (in some cases significantly) slower than the original LPA. A GPU-accelerated parallel implementation of the original LPA is available that is able to deal with large-scale datasets that do not fit into GPU memory \cite{com-kozawa17}.

%% file: 03-evaluation.tex
In this section, we first describe our experimental setup, such as the system we use and our dataset. We then investigate various static community detection algorithms, namely, the RAK algorithm, the COPRA algorithm, and the SLPA algorithm. Here we take note of the subtle details and explore various optimizations of each of the algorithms, while implementing their single-threaded and multi-threaded OpenMP-based versions.

\subsection{Experimental setup}

\subsubsection{System used}

For our experiments, we use a system consisting of two Intel Xeon Silver 4116 64-bit @ 2.10 GHz Server CPUs, and 128GB DDR4 Synchronous Registered DRAM @ 2666 MHz. Each CPU has 12 x86 cores (with 2 hyper-threads per core) and 16.5M L3 cache. Our server is running CentOS version 7.9. We use GCC version 9.3 and OpenMP version 5.0 to compile with optimization level 3 (-O3). For all experiments we keep simultaneous multi-threading (SMT) enabled.

\subsubsection{Dataset}

The graphs we use in our experiments are shown in Table \ref{tab:dataset}. All of them are obtained from the SuiteSparse Matrix Collection \cite{suite19}. The total number of vertices in the graphs vary from $74.9$ thousand to $12.0$  million, and the total number of edges vary from $811$ thousand to $304$ million. We consider all edges to be undirected and weighted with a default weight of one, and add self-loops to each vertex in all the graphs.

\input{src/tab-dataset}

\subsection{RAK algorithm}

\textbf{Raghavan Albert Kumara (RAK)}, also known as \textbf{LPA}, is a popular \textit{label-propagation} based community detection algorithm. Here, every vertex is initialized with a unique label and at every step each vertex adopts the label that most of its neighbors currently have. In the weighted version of this algorithm (which we use), each vertex adopts the label with the most interconnecting weight. Through this iterative process densely connected groups of vertices form a consensus on a unique label to form communities.

When there exist multiple communities (labels) connected to a vertex with maximum weight, we randomly pick one of them (\textbf{non-strict} approach), or pick only the first of them (\textbf{strict} approach). The algorithm converges when at least $100-n\%$ of vertices don't change their community membership (\verb|tolerance| parameter). We conduct experiments to obtain a \verb|tolerance| parameter value that would be suitable on average for most graphs. In addition, we are interested in comparing the performance of both the \textbf{non-strict} and the \textbf{strict} approaches, in terms of modularity (quality) and computation time.

For the first experiment we implement a \textbf{single-threaded CPU-based version} of the RAK algorithm \footnote{Single-threaded RAK, https://github.com/ionicf/rak-communities-seq}. We adjust \verb|tolerance| for both the \textbf{non-strict} and \textbf{strict} approaches from $0.1$ to $0.0001$. From the results, we make the following observations. \textbf{Non-strict} approach generally \textit{achieves better modularity}, with certain exceptions (\verb|coAuthors*|, \textit{road networks}). We think this has to do with graph structure. It seems to us that a \verb|tolerance| of $0.05$ would be a good choice.

For the second experiment, we implement a \textbf{multi-threaded OpenMP-based version} of the RAK algorithm \footnote{Multi-threaded OpenMP-based RAK, https://github.com/ionicf/rak-communities-openmp}. Each thread is given a separate hashtable, which it can use for choosing the most popular label among its neighbors (by weight). We initially packed the hashtables (for each thread) contiguously on a vector. Nevertheless, we observe that \textit{allocating them separately} gives us an almost \textit{2x performance improvement} over the contiguous allocation version (by using pointers to vectors). We use an OpenMP schedule of ``\verb|auto|'' and a total of $12$ threads for the time being. For optimal performance, a better schedule can be chosen through experimentation/profiling, and up to $N-1$ threads put in use (where $N$ is the number of cores/threads). Similar to the previous experiment, we adjust \verb|tolerance| from $0.1$ to $0.0001$ and compare the \textbf{sequential} and \textbf{OpenMP} implementations (\textbf{non-strict}, \textbf{strict} approaches).

From the results of the second experiment, we observe that \textbf{OpenMP-based strict RAK} performs better on average, both in terms of time and modularity. In addition, we generally see it achieve better modularity than sequential approaches (this could be due to better randomization). For a \verb|tolerance| of $0.05$, \textit{OpenMP-based strict RAK (using 12 threads)} is \textit{6.75x} faster than \textit{sequential non-strict RAK}. However, OpenMP implementations do not achieve better modularity on \verb|coAuthor*| graphs. On \textit{social networks}, \textit{OpenMP-based non-strict RAK} achieves better modularity than the \textit{strict} version. Similar to sequential RAK, we find a \verb|tolerance| of $0.05$ to be a good choice. You can find code and results for these experiments at our repository.

\subsection{COPRA algorithm}

\textbf{Community Overlap PRopagation Algorithm (COPRA)} is another \textit{label-propagation} based community detection algorithm. Unlike \textbf{RAK}, this algorithm uses \textit{multiple labels per vertex}, with each label having an associated \textit{belonging coefficient} (which sums to $1$). The algorithm is as follows. Each vertex initializes as its own community (with \textit{belonging} equal to $1$). At the start of each iteration, each vertex collects labels from its neighborhood. Here, we exclude each vertex's own labels, although not explicitly mentioned in the original paper. The collected labels are then scaled by interconnecting edge weights for weighted graphs. Labels above a certain threshold are only picked. This threshold is inversely proportional to the maximum number of labels allowed (\verb|max_labels| parameter). If all labels are below threshold, a random best label is picked. We make a vertex join its own community if it has no labels. Selected labels are then normalized such that the \textit{belonging coefficient} sums to $1$. This process is repeated until convergence.

The authors of this paper suggest using a \textit{minimum vertices per community count} to detect convergence. Alas, we do not find it to be helpful. Instead, we use a convergence condition similar to that used in \textbf{RAK}, i.e., to count the number of vertices that change their best label. Once this count falls below a certain fraction (\verb|tolerance| parameter), we consider the algorithm to have converged. The authors also mention using a synchronous version of the algorithm, where labels of each vertex are dependent only upon labels in previous iteration. Nonetheless, we find an \textbf{asynchronous approach} to converge faster (labels of each vertex can be dependent upon labels of its neighbors in the current iteration). Since our focus is on finding \textit{disjoint communities}, we consider the \textit{best label} of each vertex as the final result.

We conduct experiments to obtain a \verb|tolerance| parameter value that would be suitable on average for most graphs, both in terms of modularity and time. In addition, we adjust the \verb|max_labels| per vertex which in turn controls the \textit{threshold weight} above which labels are picked.

For the first experiment we implement a \textbf{single-threaded CPU-based version} of the COPRA algorithm \footnote{Single-threaded COPRA, https://github.com/ionicf/copra-communities-seq}. We adjust \verb|tolerance| from $0.1$ to $0.0001$, and \verb|max_labels| from $1$ to $32$. From the results, we make the following observations. On average, it seems that \textit{using a single label} is best in terms of time as well as modularity. This is indeed true in case of \textit{road networks}, but does not apply to the other classes of graphs (e.g. \textit{web graphs}, \textit{social networks}, and \textit{collaboration networks}). For example, \textit{web graphs} such as \verb|web-Stanford| and \verb|web-BerkStan| achieve best modularity with \verb|max_labels| of $8$, \verb|web-Google| attains best modularity with \verb|max_labels| of $32$, and \verb|web-NotreDame| does best with \verb|max_labels| of $16$. \verb|max_labels| of $4$-$16$ would be a good choice for such graphs. In addition it seems that on average, making the \verb|tolerance| tighter than $0.01$ has no beneficial effect on modularity. Having said that, tightening \verb|tolerance| (to any value lower than $0.1$) does not help with graphs such as \verb|web-NotreDame|, \verb|coAuthorsDBLP|, and \textit{social networks}. In any case, a \verb|tolerance| of $0.01$ would be a good choice on average.

For the second experiment, we implement a \textbf{multi-threaded OpenMP-based version} of the COPRA algorithm \footnote{Multi-threaded OpenMP-based COPRA, https://github.com/ionicf/copra-communities-openmp}. Similar to RAK, each thread is given a \textit{separate hashtable}, which it can use for choosing a set of labels from its neighbors above a certain threshold (by weight). If no label is above threshold, we pick only the best weighted label, and if not we simply join our own community. Hashtables are \textit{allocated separately} for better performance (instead of storing them contiguously on a vector). Again, we use an OpenMP schedule of ``\verb|auto|'' and a total of $12$ threads for the time being (not an optimal choice). Similar to the previous experiment, we adjust \verb|tolerance| from $0.1$ to $0.0001$, and \verb|max_labels| from $1$ to $32$. From the results, we observe that \textbf{OpenMP-based approaches} are \textbf{faster} than \textit{sequential}, and also seem to achieve better modularity. The rest of our observations are similar to that of the first experiment.

\subsection{SLPA algorithm}

\textbf{Speaker-listener Label Propagation Algorithm (SLPA)} is yet another \textit{label-propagation} based community detection algorithm. Here, each vertex has a fixed label memory where it remembers all the popular labels that it has listened to. The size of this memory is fixed to the total number of iterations to be performed \textit{plus one}. The algorithm is as follows. Each vertex is first initialized such that it remembers itself as popular. At the start of each iteration, each neighbor (of a vertex) speaks one of the random labels in its memory to the vertex. The vertex (listener) adds the most popular label to its memory. This is repeated until a fixed number of iterations is performed, equal to \verb|memory_size| minus one. In order to accelerate the algorithm, we allow early convergence if at least $100-n\%$ of vertices memorize their previous label (\verb|tolerance| parameter). We use a \verb|tolerance| of $0.05$. As we are interested in finding \textit{disjoint communities}, we pick the most popular label in the memory of each vertex as its community.

Unlike \textbf{RAK} and \textbf{COPRA}, this is a \textit{randomized historical-label} based technique. A \textit{random number generator (RNG)} is used in order for each neighbor of a given vertex to pick a random label to speak from its memory. We originally used C++'s \textit{default random engine} (probably \textit{Mersenne twister}) which seemed slow. We updated it to use a \textit{xorshift32 random engine}. The listener vertex listens to all of its neighbors, and based on edge weight, picks the most popular label. In \textbf{strict} mode, the listener picks the first most popular label to record in its memory, and in \textbf{non-strict} mode, the listener picks randomly one of the most popular labels to record. It records this in the next index in its memory.

We conduct an experiment, where we adjust the \verb|memory_size| per vertex which in turn controls the \textit{maximum number of iterations} needed to be performed \footnote{Single-threaded SLPA, https://github.com/ionicf/slpa-communities-seq}. This is done in order to obtain a \verb|memory_size| parameter value that would be suitable on average for most graphs, both in terms of modularity and time.

From the results, we observe that increasing the \verb|memory_size| per vertex increases the time required for completion, while also increasing the final modularity. Anyhow, on graphs \verb|soc-Slashdot*| modularity seems to decrease with increasing \verb|memory_size|. We also observe that the \textbf{strict} variant is slightly faster than the \textbf{non-strict} one, while achieving similar modularity. On \textit{road networks}, the \textbf{strict} approach seems to obtain lower modularity, while on other classes of graphs, the \textbf{strict} approach achieves higher modularity. However, it seems that the SLPA algorithm does not provide a good return on investment (it takes longer but does not achieve great modularity). It should be noted that we do not perform any post-processing of communities (such as splitting disconnected communities).

%% file: src/tab-dataset.tex
\begin{table}[!ht]
\centering
\caption{List of 17 graphs used in our experiments. All edges of each graph were duplicated in the reverse direction, thus making them undirected, and the weight of each edge was set to 1. Here, $|V|$ is the total number of vertices, $|E|$ is the total number of edges (after making the graph undirected), and $D_{avg}$ is the average degree of vertices. The number of vertices and edges are rounded to the nearest thousand (K) or million (M), as appropriate.}
\label{tab:dataset}
\begin{tabular}{||c||c|c|c|c||}
  \toprule
  \textbf{Graph} &
  \textbf{$|V|$} &
  \textbf{$|E|$} &
  \textbf{$D_{avg}$} \\
  \midrule
  \multicolumn{4}{|c|}{Web Graphs} \\ \hline
    web-Stanford & 282K & 3.99M & 14.1 \\ \hline
    web-BerkStan & 685K & 13.3M & 19.4 \\ \hline
    web-Google & 916K & 8.64M & 9.43 \\ \hline
    web-NotreDame & 326K & 2.21M & 6.78 \\ \hline
    indochina-2004 & 7.41M & 304M & 41.0 \\ \hline
    \multicolumn{4}{|c|}{Social Networks} \\ \hline
    soc-Slashdot0811 & 77.4K & 1.02M & 13.2 \\ \hline
    soc-Slashdot0902 & 82.2K & 1.09M & 13.3 \\ \hline
    soc-Epinions1 & 75.9K & 811K & 10.7 \\ \hline
    soc-LiveJournal1 & 4.85M & 86.2M & 17.8 \\ \hline
    \multicolumn{4}{|c|}{Collaboration Networks} \\ \hline
    coAuthorsDBLP & 299K & 1.96M & 6.56 \\ \hline
    coAuthorsCiteseer & 227K & 1.63M & 7.18 \\ \hline
    coPapersCiteseer & 434K & 32.1M & 74.0 \\ \hline
    coPapersDBLP & 540K & 30.5M & 56.5 \\ \hline
    \multicolumn{4}{|c|}{Road Networks} \\ \hline
    italy\_osm & 6.69M & 14.0M & 2.09 \\ \hline
    great-britain\_osm & 7.73M & 16.3M & 2.11 \\ \hline
    germany\_osm & 11.5M & 24.7M & 2.15 \\ \hline
    asia\_osm & 12.0M & 25.4M & 2.12 \\ \hline
  \bottomrule
\end{tabular}
\end{table}

%% file: 04-conclusion.tex
We examine a number of label-propagation-based static community detection algorithms, including the RAK, COPRA, and SLPA. While implementing their single-threaded and multi-threaded OpenMP-based versions, we pay attention to the minute details of each algorithm, make any useful tweaks/optimizations, and obtain suitable parameter values. The \textbf{RAK} algorithm performs well with a \verb|tolerance| of $0.05$ (which determines the maximum percentage of vertices that change their community membership), with \textit{strict RAK} performing better than non-strict version in parallel implementation (except on social networks), and \textit{OpenMP-based strict RAK} with 12 threads was \textit{6.75x} faster than the sequential non-strict RAK. The \textbf{COPRA} algorithm works well with a single label for \textit{road networks}, and \verb|max_labels| of $4$-$16$ for \textit{other classes} of graphs. We use an asynchronous version of the algorithm (which converges faster), and use the same convergence condition as the RAK algorithm. A \verb|tolerance| of $0.01$ is a good choice on average, and OpenMP-based implementation is faster and achieves \textit{better modularity} than sequential. For the SLPA algorithm, we use a xorshift32 random engine for generation of random numbers, and again use the same convergence condition as the RAK algorithm. The strict variant of SLPA performs better than the non-strict one (except on road networks), performs well with increasing \verb|memory_size|, but overall doesn't offer a favourable return on investment (its slow). We therefore recommend the reader to stick to the \textit{RAK algorithm} for label-propagation based disjoint community detection.